\documentclass[10pt]{iopart}
\usepackage{epsfig,amsfonts,cite}
\usepackage{pstricks}

\def\be{\begin{equation}}
\def\ee{\end{equation}}
\def\bea{\begin{eqnarray}}
\def\eea{\end{eqnarray}}

\def\Tr{\mathop{\rm  Tr}}
\def\bsigma{\mbox{\boldmath $\sigma$}}

\def\bphi{\mbox{\boldmath $\phi$}}

\def\bTheta{\mbox{\boldmath $\Theta$}}

\def\mat#1{\hat{\hbox{{\sffamily \bfseries #1}}}}

\def\<{\langle}
\def\>{\rangle}

\begin{document}
\jl{31}
\title{\sffamily \bfseries 
	Introduction to thermodynamics of spin models in the Hamiltonian limit}

\author{Bertrand Berche$^{(a,b)}$ and Alexander L\'opez$^{(b)}$}
\address{$^{(a)}$Groupe M, 
Laboratoire de Physique des Mat\'eriaux, UMR  CNRS No 7556,\\ 
Universit\'e Henri
Poincar\'e, Nancy 1,\\
B.P. 239, F-54506 Vand\oe uvre les Nancy, France\\
$^{(b)}$Instituto Venezolano de Investigaciones Cient\'\i ficas, \\
Centro de F\'\i sica, \\
Carr. Panamericana, km 11,
Altos de Pipe, \\
Aptdo 21827, \\
1020-A Caracas, Venezuela 
}
\date{\today}
\begin{abstract}
\baselineskip=9pt
A didactic description of the thermodynamic properties of classical spin
systems is given in terms of their quantum counterpart in the Hamiltonian
limit. Emphasis is on 
the construction of the relevant Hamiltonian, and the calculation of
thermal averages is explicitly done in the case of small systems
described, in Hamiltonian field theory, by small matrices.
The targeted students are those of a graduate statistical physics course
\\
{\bf R\'esum\'e.} 
On donne une description didactique des propri\'et\'es thermodynamiques
de syst\`emes de spins classiques en termes de leur 
\'equivalent quantique dans la limite hamiltonienne. On insiste sur
la construction de l'hamiltonien et le calcul des moyennes thermiques
est effectu\'e explicitement dans le cas de syst\`emes de petite
taille, d\'ecrits dans la th\'eorie hamiltonienne des champs par des
petites matrices. Le public vis\'e est celui d'\'etudiants suivant un
cours avanc\'es en 
physique statistique.
\end{abstract}
\pacs{05.50.+q Lattice theory and statistics (Ising, Potts, etc.),\\
\phantom{PACS numbers: }64.60.Fr Equilibrium properties near critical 
points, critical\\
\phantom{PACS numbers: \qquad}exponents,\\
\phantom{PACS numbers: }75.10.Hk Classical spin models}
\ \hskip2.37cm\today



\section{Introduction}
\label{sec:intro}
A rather interesting method for studying thermal fluctuations of 
{\em $(d+1)$-dimensional}
interacting {\em classical} degrees of freedom is provided through the 
corresponding analysis of quantum fluctuations in a 
{\em $d$-dimensional} system of
interacting {\em quantum} degrees of freedom.
The correspondence is detailed in the remarkable review 
of Kogut~\cite{Kogut79}.
It can be understood as a re-wording of the transfer matrix 
formalism~\cite{SchultzLiebMattis64}
of classical systems in a special limit where the transfer matrix takes
a simplified form, the so-called Hamiltonian limit~\cite{FradkinSusskind78}.
At the origin of the correspondence, there is the transition amplitude between
quantum states in the path integral formulation,
\be
	{\tt Amplitude} = \sum_{\tt paths}\exp\frac i\hbar S[x(t)]
	\label{eqAmpl}
\ee
where the classical action is a functional of $x(t)$,  
the time integral of a Lagrangian
$S=\int_{t_a}^{t_b} L(x,\dot x) d t$. 
Here we consider  the transition amplitude
of a point particle (space dimensionality of the quantum system $d=0$) 
for the sake of simplicity.
Equation~(\ref{eqAmpl}) looks a bit similar to the partition function
of {\em something} to be specified,
\be
	{\tt P.F.} = \sum_{\tt configurations}\exp (-\beta E\{{\tt d.o.f.}\}),
\ee
provided that we change the imaginary argument of the exponential to a real
one and we give sense to the degrees of freedom (d.o.f.)
and to the sum over their configurations. 
This transformation is achieved through the definition of an
imaginary time, also called Euclidean time, 
\be t=-i\tau\label{eqImaginaryTime}\ee
in terms of which $L(x,\dot x)=-H(x,p)$, $p=mx'$, $\dot x=dx/dt$ and
$x'=dx/d\tau$. The phase factor of equation~(\ref{eqAmpl}) becomes,
with equation~(\ref{eqImaginaryTime}),
\be
	\frac i\hbar S[x(t)]=-\frac 1\hbar\int_{\tau_a}^{\tau_b} H(x,p)d\tau
\ee
where $H(x,p)=\frac{p^2}{2m}+V(x)$.
Equation~(\ref{eqAmpl}) has a mathematical sense through discretization
of the time axis, $t_{i+1}-t_i=\delta=-i\epsilon$, $\epsilon\to 0$, 
$t_b-t_a=N\delta$, 
$x_i=x(t_i)$.
The action becomes~\footnote{We denote by $S[x(\tau)]$ the functional of a 
continuous function and $\sum_iH\{x_i\}$
the corresponding sum over a set of discrete variables.}
\be
S[x(\tau)]\leadsto\epsilon\sum_{i=1}^{N-1}\left(\frac 12m
\Bigl(\frac{x_{i+1}-x_i}{\epsilon}\Bigr)^2+V(x_i)\right)
\ee
and the transition
amplitude eventually reads as
the partition function of a set of $N$ classical
d.o.f. $\{x_i\}$ on a 1-dimensional lattice
\bea
	{\tt Amplitude} &=& \int\Bigl(\prod_i dx_i\Bigr) 
	\exp (-\hbar^{-1} \sum_{i} H\{ x_i\})
	\nonumber\\
	{\tt P.F.} &=& \sum_{\{x_i\}}\exp (-\beta \sum_i H\{x_i\}).
\label{eq-correspondence}
\eea
$H\{x_i\}$ is a classical  energy density depending on the set of values
taken by the d.o.f., ($-\infty<x_i<+\infty$ if there is no 
restriction specified in the original quantum problem)  
and leads after summation
to the energy of a configuration appearing in the Boltzmann weight.
The role of $\beta$ (thermal fluctuations) is played by $\hbar^{-1}$
(quantum fluctuations). The classical limit $\hbar \to 0$ where quantum
fluctuations are suppressed corresponds to $\beta\to\infty$ in the classical
system, i.e. suppression of thermal fluctuations.
The correspondence (\ref{eq-correspondence}) is more generally valid than in 
the simple case of a point particle. To the statistical physics problem 
of classical degrees of freedom living in $d+1$ space dimensions, 
there corresponds a quantum problem in $d$ dimensions where the fluctuations
result from competition between non commutating variables. 
The partition function
of the former problem involves quantum transitions between multi-particle
states in the latter formulation and the Hamiltonian limit is the simplest
formulation of the quantum problem when only survive transitions between
single-particle quantum states.

This approach deserves some attention and might be taught in graduate
statistical physics courses. In the following, we remind how
the thermodynamic 
properties of a classical system might be obtained from their quantum
counterpart, then we apply the technique to an approximate determination of the
critical properties of two-dimensional classical spin
models, namely the Ising model and the $3-$ and $4-$state Potts models.

\section{Thermodynamics}
In the quantum version of the problem, 
we may define a time evolution
operator in terms of which the Feynman kernel is 
\bea
	K( x_b,t_b| x_a,t_a)&=&\< x_b|
        \exp (- \frac i\hbar \mat{H} (t_b-t_a))| x_a\>
	\nonumber\\
	&=&\int \< x_b|\mat{T}| x_{N-1}\>d  x_{N-1}
	\< x_{N-1}|\mat{T}| x_{N-2}\>
	d  x_{N-2}
	\< x_{N-2}|\dots\nonumber\\
	&\phantom{=}&\qquad\qquad\qquad\dots d x_2 \< x_2|\mat{T}| x_{1}\>
	d  x_{1}\< x_1|\mat{T}| x_{a}\>
	\nonumber\\
	&=&\< x_b|\mat{T}^N| x_a\>,
\eea
where $\mat{T}$ is the infinitesimal time evolution operator which, in the
Euclidean time, becomes the transfer matrix 
$\mat{T}=\exp(-\epsilon\mat{H}/\hbar)$,
$\delta=-i\epsilon$. $N\epsilon$ is the length of the system in the
supplementary time direction (see Fig.~\ref{Fig0}).
Summing over initial states, when periodic boundary conditions in the
time direction are imposed, $ x_b= x_a= x_0$, we get the partition function 
from
\be
	Z=\int d x_0K( x_0,-iN\epsilon| x_0,0)=\Tr\mat{T}^N.
\ee
The thermodynamic limit $N\to\infty$ ensures projection onto the 
{\em ground state},
\bea
	\mat{T}^N&=&t_0^N\Big[|0\>\<0|+\sum_{\alpha\not = 0}|\alpha\>
	(t_\alpha/t_0)^N\<\alpha|\Big]
	\to |0\>t_0^N\<0|\quad t_0>t_1>\dots 
\eea
The partition function is thus 
determined by the largest eigenvalue $t_0$ of the 
transfer matrix,
\be
	Z\to t_0^N=\e^{-N\epsilon E_0/\hbar},
\ee
where $E_0$ is the ground state energy of the quantum Hamiltonian $\mat{H}$.
The free energy density (i.e. per time slice) follows 
\be
	f=-\lim_{N\to\infty}(N\epsilon)^{-1}\hbar\ln  Z\to  E_0.
\ee

\begin{figure} [th]
\vspace{-1cm}
        \epsfxsize=7.0cm
        \begin{center}
        \mbox{\epsfbox{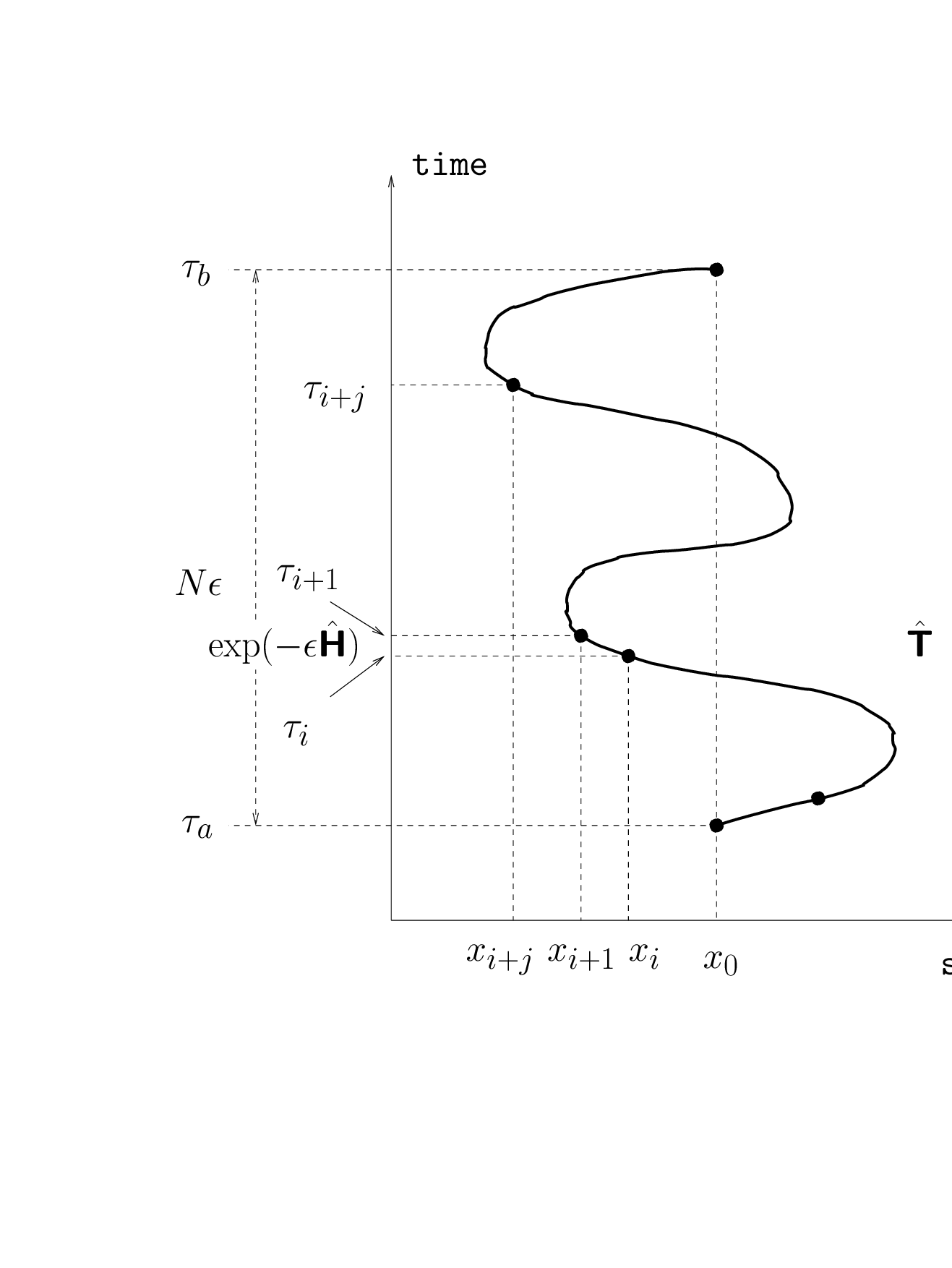}\qquad}
        \end{center}\vskip -2.5cm
        \caption{Space-time quantum path and the corresponding classical
	chain of {\tt d.o.f} $x_i$. Periodic boundary conditions in
	time direction are assumed.}
        \label{Fig0}  
\end{figure}

The (time) correlation function of some local quantities $\phi_{x}$ 
depending on the classical d.o.f. $x$, 
$\<\phi_{ x_i}\phi_{ x_{i+j}}\>$, is expressed in 
terms of 
the transfer matrix through

\bea
	\<\phi_{ x_i}\phi_{ x_{i+j}}\>&=&\lim_{N\to\infty}
	Z^{-1}\sum_{\{ x_i\}}\phi_{ x_i}\phi_{ x_{i+j}}
	\e^{-\beta E\{ x_i\}}\nonumber\\
	&=&\lim_{N\to\infty}Z^{-1}\int 
	\< x_0|\mat{T}| x_{N-1}\>d  x_{N-1}\< x_{N-1}| 
	\dots\nonumber\\
	&\phantom{=}&\qquad
	\dots \mat T| x_{i+j}\>\phi_{ x_{i+j}}d x_{i+j}\< x_{i+j}|\dots
	\mat T| x_{i}\>\phi_{ x_i}d x_{i}\< x_{i}|\dots\nonumber\\
	&\phantom{=}&\qquad\dots 
	\mat T| x_{1}\>d  x_{1}\< x_1|\mat{T}| x_{0}\>\nonumber\\
	&=&\lim_{N\to\infty}
	\frac{\Tr (\mat{T}^i\hat{\bphi}\mat{T}^{j}\hat{\bphi}\mat{T}^{N-i-j})}
	{\Tr \mat{T}^N},\label{eq-11}
\eea
where diagonal operators in the $\{| x_k\>\}$ basis have been introduced,
\be
	\hat{\bphi}=\int d x_k | x_k\>\phi_{ x_k}\< x_k|.
\ee

Using the eigenstates of the transfer matrix, the correlation function
(\ref{eq-11}) becomes
\be
	\<\phi_{ x_i}\phi_{ x_{i+j}}\>=|\<0|\hat{\bphi}|0\>|^2+\sum_{\alpha\not = 0}
		|\<0|\hat{\bphi}|\alpha\>|^2
	\left(\frac{t_\alpha}{t_0}\right)^j.
\ee
The connected part of the correlation function,  $G_\phi(j)$, 
is obtained after subtraction of the
ground state expectation value, 
and in the thermodynamic limit it is dominated by the
first eigenstate $|{\beta_\phi}\>$ such that the matrix element of $\hat{\bphi}$
between this state and the ground state does not vanish,
\be
	G_\phi(j)\to |\<0|\hat{\bphi}|{\beta_\phi}\>|^2
	\left(\frac{t_\beta}{t_0}\right)^j=	
	|\<0|\hat{\bphi}|{\beta_\phi}\>|^2
	\e^{-j\epsilon\hbar^{-1}(E_\beta-E_0)}.
\ee
The matrix element in prefactor measures the average of the field squared,
\be
	\<{\tt average\  field}\>=|\<0|\hat{\bphi}|{\beta_\phi}\>|,
\label{eq-elmat}\ee
and the exponential decay along the time direction
allows to define the correlation length in terms of an inverse gap
($\hbar =1$),
\be
	\frac{1}{\xi_\phi}=E_\beta-E_0={\tt gap}_\phi.
\ee

A challenging problem in critical phenomena is the identification of the
universality class of a given model. Due to scaling relations among
critical exponents, the knowledge of two of them determines the whole
set of exponents (see table~\ref{tabNomenclatura}).
From finite size-scaling (FSS) behaviour of the critical 
densities~\cite{Barber83}, we expect
power law behaviours of local quantities in terms of the  finite size
$L$ in the space direction,
\bea
&&\<{\tt Energy\ density}\>\sim L^{-x_\epsilon},\label{eq-e}\\
&&\<{\tt Order\ parameter}\> \sim L^{-x_\sigma}.\label{eq-m}
\eea
These relations may be taken as definitions of 
the scaling
dimensions of the energy and order parameter density,
$x_\epsilon=(1-\alpha)/\nu$ and $x_\sigma=\beta/\nu$. The commonly accepted 
notation for the critical exponents is reminded in table~\ref{tabNomenclatura}.
\begin{table}
        \caption{Usual definitions of the critical exponents. The 
	physical quantities
	are referred to as in the context of 
	a magnetic system.}
        \label{tabNomenclatura}
	\begin{indented}
	\item[]
        \begin{tabular}{llllll}
        \br
Quantity & leading singularity & condition & scaling relation\\
        \mr
Specific heat & $C(T)\sim |T-T_c|^{-\alpha}$ & $h=0$ & 
	$\frac {1-\alpha}{\nu}=x_\epsilon$\\
Magnetization & $M(T)\sim |T-T_c|^{\beta}$ &  $h=0,\ T<T_c$
	& $\beta/\nu=x_\sigma$\\
Susceptibility & $\chi(T)\sim |T-T_c|^{-\gamma}$ & $h=0$
	& $\gamma/\nu=2-2x_\sigma$\\
Critical isotherm & $M(h)\sim |h|^{1/\delta}$ & $T=T_c$
	& $\delta = \frac{d-x_\sigma}{x_\sigma}$\\
Correlation length & $\xi(T)\sim |T-T_c|^{-\nu}$ & $h=0$
	& $\nu=(d-x_\epsilon)^{-1}$\\
Correlation function & $G(r)\sim r^{-(d-2+\eta)}$ & $h=0,\ T=T_c$
	& $\eta=2-d+2x$\\
        \br
        \end{tabular}
	\end{indented}
\end{table}
Imagine that we consider a more general quantum problem in 
$d-1$-space dimensions. 
The classical counterpart is defined in $d$-dimensions
(space {\em and} time) and 
the susceptibility for example follows from the integration of the
correlation function over these $d$ dimensions. Let us write schematically
what happens in
the case of a translation invariant (in $d-1$ dimensions) matrix element,
\be G_\phi({\tt time})=|\<0|\hat{\bphi}|{\beta_\phi}\>|^2
	\e^{-{\tt gap}_\phi\times{\tt time}},\label{eq-corr}
\ee
\bea	\chi&=&\sum_{{\tt space}}\int G_\phi({\tt time})
	\ \!d\ \!{\tt time} \nonumber\\
	&\simeq& 
	L^{d-1} |\<0|\hat{\bphi}|{\beta_\phi}\>|^2 \frac{1}{{\tt gap}_\phi}.
	\label{eq-chi}
\eea
The matrix element $|\<0|\hat{\bphi}|{\beta_\phi}\>|$ should scale like
$L^{-x_\phi}$ (see e.g. equations~(\ref{eq-e}) and (\ref{eq-m})), 
hence relation~(\ref{eq-chi}) requires that the gap scales
according to ${\tt gap}_\phi\sim L^{-1}$ in order to restore the usual
scaling of the susceptibility $\chi\sim L^{\gamma/\nu}$ with 
$\gamma/\nu=2-\eta=d-2x_\phi$ (see table~\ref{tabNomenclatura}). 
Note that the gap being an inverse 
correlation length, its scaling inversely proportional to the typical 
linear size of the system logically means that the correlation length
at criticality is locked at that size.

The scaling of the gap is more constrained in $2d$ where rather
powerful techniques apply.
Conformal invariance indeed provides quite efficient methods for the 
determination
of critical exponents of two-dimensional critical systems~\cite{Cardy86}. 
The cylinder geometry is relevant in the study of quantum chains, since
such a chain with periodic boundary conditions in the space direction just
corresponds to an infinitely long classical cylinder 
($1+1$ dimensions)
of complex
coordinates $w={\tt time}+i\times{\tt space}$ (here, Euclidean time is 
assumed). This former 
geometry follows from the infinite two-dimensional plane $z=r\e^{i\theta}$ 
through the standard logarithmic mapping $w(z)=\frac L{2\pi}\ln z$ and,
at the critical point, the correlation functions transform according
to conformal covariance, leading along the cylinder to~\cite{Cardy84}
\be G_\phi({\tt time})=\left(\frac{2\pi}{L}\right)^{2x_\phi}
	\e^{-\frac{2\pi}{L}x_\phi\times{\tt time}}.\label{eqCorrConf}
\ee
From comparison with equation~(\ref{eq-corr}),
the critical correlation length amplitude appears
universal and its value related to the corresponding critical exponent
through the simple relation
\be {\tt gap}_\phi=\frac{2\pi}{L}x_\phi.\label{eq-gap}\ee
The matrix elements are also predicted by conformal invariance and follow
from equation~(\ref{eqCorrConf}),
\be
|\<0|\hat{\bphi}|\beta\>|=\left(\frac{2\pi}{L}\right)^{x_\phi}.
\ee

A prescription before using any conformal invariance result concerns the
scaling of the whole spectrum. Multiplying the Hamiltonian by an arbitrary
number of course changes the scale of the spectrum. Gap scaling from
equation~(\ref{eq-gap}) requires to fix the normalization in such a way that
the sound velocity is unity~\cite{GehlenEtAl85}. 
The sound velocity might be defined in the
long-wavelength limit, ${\tt sound\ velocity}=\Delta E/\Delta k$, where
$\Delta E$ is for instance measured by gaps in the bottom of the spectrum
and $\Delta k=2\pi/L$ is given by the quantization step of wave vectors.

\section{Quantum Ising chain}
The two-dimensional Ising model is often used as the paradigmatic
illustration of second-order phase transitions. It is one of the most simple
non trivial models and in the following we will show how its quantum 
counterpart, the quantum Ising chain in a transverse field, is built.
\subsection{The Hamiltonian limit}
\begin{figure} [th]
\vspace{-0.5cm}
        \epsfxsize=7.5cm
        \begin{center}
        \mbox{\epsfbox{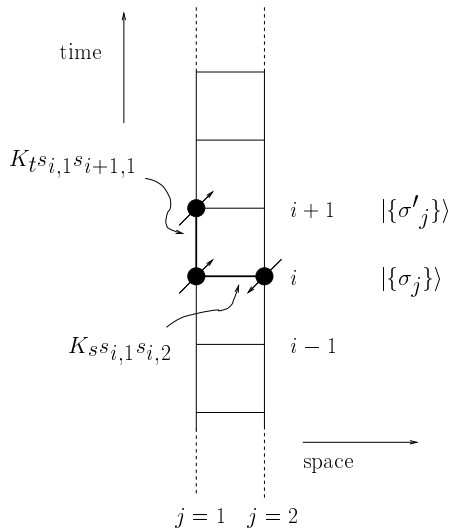}\qquad}
        \end{center}\vskip -2cm
        \caption{Infinite ladder of Ising spins.}
        \label{Fig1}  
\end{figure}

Let us first consider a ladder of classical Ising spins 
${s}_{i,j}=\pm 1$,
$-\infty < i< +\infty$, and $j=1,2$ (see figure~\ref{Fig1}). 
The nearest neighbour interactions
are denoted as $K_s=\beta J_s$ in the space ($j$) direction and
$K_t=\beta J_t$ in the time ($i$) direction. 
The energy of a configuration is a sum over time slices,
\be
	-\beta E\{{s}_{i,j}\}=\sum_i(K_s{s}_{i,1}{s}_{i,2}+K_t(
	{s}_{i,1}{s}_{i+1,1}+{s}_{i,2}{s}_{i+1,2})),
\ee
and the partition function reads as
\bea
	Z&=&\Tr_{\{{s}_{i,j}\}}\prod_i \e^{K_s{s}_{i,1}{s}_{i,2}+K_t(
	{s}_{i,1}{s}_{i+1,1}+{s}_{i,2}{s}_{i+1,2})}
	=\Tr_{\{{s}_{i,j}\}}\prod_i T_{i,i+1},
\eea
where  $T_{i,i+1}$ is an element of some transfer matrix between
row states $|\{\sigma_j\}\>=|s_{i,1},s_{i,2}\>$ 
and $|\{{\sigma'}_j\}\>=|s_{i+1,1},s_{i+1,2}\>$ 
at rows $i$ and $i+1$ (see figure~\ref{Fig1}),
\be
T_{i,i+1}=\<\{\sigma'_j\}|\mat{T}|\{\sigma_j\}\>.
\ee 
$\mat{T}$ is a $4\times 4$ matrix, 
since the two spins $\sigma_1,\sigma_2$ have four different  
configurations.
The space coupling term is diagonal, 
$\delta_{\{\sigma_j\},\{\sigma'_j\}}\times\exp(K_s\sigma_{1}\sigma_{2})$,

\be
\begin{array}{c|c}
\begin{pspicture}(0,0)(0.5,0.5)
\psline(0.5,0)(0,0.5)
\rput(-0.07,0.0){$\sigma'_1,\sigma'_2$}
\rput(0.58,0.47){$\sigma_1,\sigma_2$}
\end{pspicture}
 & 
\begin{array}{cccc}
|\uparrow,\uparrow\> & |\uparrow,\downarrow\> & |\downarrow,\uparrow\> & 
|\downarrow,\downarrow\> \\
\end{array} 
\\ \noalign{\vskip3pt} \hline \noalign{\vskip2pt}
\begin{array}{c}
\<\uparrow,\uparrow| \\ \<\uparrow,\downarrow| \\ \<\downarrow,\uparrow| \\ 
\<\downarrow,\downarrow|
\end{array} &
\left(
\begin{array}{cccc}
\e^{K_s} & 0 & 0 & 0 \\ 
0 & \e^{-K_s} & 0 & 0 \\
0 &  0 & \e^{-K_s} & 0 \\
0 & 0 & 0 &\e^{K_s} 
\end{array} 
\right),\\
\end{array}
\label{eqSpaceMatrix}
\ee
while a non-diagonal contribution due to possible spin flips in the time
direction, $\exp[K_t(\sigma_{1}\sigma'_{1}+\sigma_{2}\sigma'_{2})]$,
leads to the following matrix

\be
\begin{array}{c|c}
\begin{pspicture}(0,0)(0.5,0.5)
\psline(0.5,0)(0,0.5)
\rput(-0.07,0.0){$\sigma'_1,\sigma'_2$}
\rput(0.58,0.47){$\sigma_1,\sigma_2$}
\end{pspicture}
 & 
\begin{array}{cccc}
|\uparrow,\uparrow\> & |\uparrow,\downarrow\> & |\downarrow,\uparrow\> & 
|\downarrow,\downarrow\> \\
\end{array} 
\\ \noalign{\vskip3pt} \hline \noalign{\vskip2pt}
\begin{array}{c}
\<\uparrow,\uparrow| \\ \<\uparrow,\downarrow| \\ \<\downarrow,\uparrow| \\ 
\<\downarrow,\downarrow|
\end{array} &
\left(
\begin{array}{cccc}
\e^{2K_t} & 1 & 1 & \e^{-2K_t} \\ 
1 & \e^{2K_t} & \e^{-2K_t} & 1 \\
1 &  \e^{-2K_t} & \e^{2K_t} & 1 \\
\e^{-2K_t} & 1 & 1 &\e^{2K_t} 
\end{array} 
\right).\\
\end{array}
\label{eqTimeMatrix}
\ee
The transfer matrix $\mat{T}$ should be identified through its 
matrix elements. For
the diagonal term, we  introduce {\em diagonal operators} in the 
$|\{\sigma_j\}\>$ basis, namely
\be
	\fl \hat{\bsigma}_z(1)=\sum_{\sigma_1,\sigma_2}
	|\sigma_1,\sigma_2\>\sigma_1\<\sigma_1,\sigma_2|,
	\quad
	[\hat{\bsigma}_z(1)]=
	\left(\begin{array}{cc}1 & 0 \\ 0 & -1\end{array}\right)_1	
	\otimes
	\left(\begin{array}{cc}1 & 0 \\ 0 & 1\end{array}\right)_2	
	,
\ee
\be	\fl\hat{\bsigma}_z(2)=\sum_{\sigma_1,\sigma_2}
	|\sigma_1,\sigma_2\>\sigma_2\<\sigma_1,\sigma_2|,
	\quad
	[\hat{\bsigma}_z(2)]=
	\left(\begin{array}{cc}1 & 0 \\ 0 & 1\end{array}\right)_1	
	\otimes
	\left(\begin{array}{cc}1 & 0 \\ 0 & -1\end{array}\right)_2	
	,
\ee
where $(\mat{1})_j$ and $(\hat{\bsigma}_z)_j$ represent the $2\times 2$
identity or Pauli
matrix acting only on variables $\sigma_j$ at site $j$.
The matrix in equation~(\ref{eqSpaceMatrix}) is identified to that of the
operator
\be
	\exp(K_s\hat{\bsigma}_z(1)\hat{\bsigma}_z(2)).
\ee
For the flipping term, we introduce {\em transition operators}
\be
	\fl \hat{\bsigma}_x(1)
	|\sigma_1,\sigma_2\>=|-\sigma_1,\sigma_2\>,
	\quad
	[\hat{\bsigma}_x(1)]=
	\left(\begin{array}{cc}0 & 1 \\ 1 & 0\end{array}\right)_1	
	\otimes
	\left(\begin{array}{cc}1 & 0 \\ 0 & 1\end{array}\right)_2	
	,
\ee
\be	
	\fl \hat{\bsigma}_x(2)
	|\sigma_1,\sigma_2\>=|\sigma_1,-\sigma_2\>,
	\quad
	[\hat{\bsigma}_x(2)]=
	\left(\begin{array}{cc}1 & 0 \\ 0 & 1\end{array}\right)_1	
	\otimes
	\left(\begin{array}{cc}0 & 1 \\ 1 & 0\end{array}\right)_2	
	,
\ee
and since $\hat{\bsigma}^2_x(j)=\mat{1}$, we have, for any
value of $K_t^\star$,  the useful identity
\be
	\exp[K_t^\star\hat{\bsigma}_x(j)]=\mat{1}\cosh K_t^\star+
	\hat{\bsigma}_x(j)\sinh K_t^\star,
\label{eqCoshSinh}
\ee
so that the matrix in equation~(\ref{eqTimeMatrix}) is identified to that of 
the operator $\e^{K_t^\star(\hat{\bsigma}_x(1)+\hat{\bsigma}_x(2))}$,
\be
\fl 
\left(
\begin{array}{cccc}
\cosh^2 K_t^\star & \cosh K_t^\star\sinh K_t^\star 
	& \cosh K_t^\star\sinh K_t^\star & \sinh^2 K_t^\star \\ 
\cosh K_t^\star\sinh K_t^\star & \cosh^2 K_t^\star 
	& \sinh^2 K_t^\star & \cosh K_t^\star\sinh K_t^\star \\
\cosh K_t^\star\sinh K_t^\star &  \sinh^2 K_t^\star 
	& \cosh^2 K_t^\star & \cosh K_t^\star\sinh K_t^\star \\
\sinh^2 K_t^\star & \cosh K_t^\star\sinh K_t^\star 
	& \cosh K_t^\star\sinh K_t^\star &\cosh^2 K_t^\star
\end{array} 
\right),
\label{eqTimeMatrixStar}
\ee
(up to a prefactor which only shifts the free energy by a constant)
provided that we demand
\be\tanh K_t^\star=\e^{-2K_t}.\ee
This relation is equivalent to the usual duality relation
\be\sinh 2K_t\sinh 2K_t^\star=1.\label{eq-duality}\ee

The transfer matrix eventually follows (we use $\e^{{\bf A}+{\bf B}}
=\e^{\bf A}\e^{\bf B}\e^{-[{\bf A},{\bf B}]/2}\dots$ and already take into 
account the simplification due to the extreme anisotropic limit which 
eliminates all correction terms)
\be
	\mat{T}=\exp(K_s\hat{\bsigma}_z(1)\hat{\bsigma}_z(2))\times
		\exp(K_t^\star(\hat{\bsigma}_x(1)+\hat{\bsigma}_x(2)).
\ee
The multiple spin flip terms appear in the expansion of the exponential
through product of terms like in equation~(\ref{eqCoshSinh}).
The generalization to a lattice of width $L$ in the space direction
(we choose now for the rest of the paper periodic boundary conditions 
$\hat{\bsigma}_{x,z}(L+1)=\hat{\bsigma}_{x,z}(1)$ in space direction)
is straightforward, summing over $j=1,L$,
\be
	\mat{T}=\exp(K_s\sum_{j=1}^L\hat{\bsigma}_z(j)
	\hat{\bsigma}_z(j+1))\times
		\exp(K_t^\star\sum_{j=1}^L\hat{\bsigma}_x(j)).
\label{eqTITF}
\ee
The transfer matrix has a complicated structure, since arbitrary large
numbers of single spin transitions may simultaneously occur and the
matrix representation of $\mat T$  is expected to be dense. 
It may be considerably
simplified in the extreme  anisotropic limit also called Hamiltonian limit,
since then only single spin transitions survive and $\mat H$ is a sparser
matrix.
Remember that the relation between the transfer matrix and the Hamiltonian
involves 
the lattice spacing $\epsilon$ in the time direction which tends 
to zero, 
$\mat{T}=\exp(-\epsilon\mat{H})=\mat{1}-\epsilon\mat{H}+O(\epsilon^2)$. 
A simple
limit of $\mat{T}$ in equation~(\ref{eqTITF}) is obtained when $K_s\to 0$,
$K_t\to\infty$ (or $K_t^\star\to 0$), with $K_s/K_t^\star=\lambda=O(1)$,
and in order to avoid unnecessary constants, we may set 
$\epsilon=2K_t^\star$~\footnote{The 
normalization with a factor 2 ensures a sound
velocity equal to 1 (i.e. a linear dispersion relation at long wavelength
with slope unity) and is necessary in order to compare later with conformal
invariance predictions (see discussion).} to get
\be
	\mat{H}=-\frac 12\lambda\sum_{j=1}^L\hat{\bsigma}_z(j)
	\hat{\bsigma}_z(j+1))
	-\frac 12\sum_{j=1}^L
	\hat{\bsigma}_x(j).
\label{eqHITF}
\ee
The critical line of the two-dimensional anisotropic 
classical system, obtained through
duality i.e. when the relation $\sinh 2K_s\sinh 2K_t=1$ is fulfilled, 
is equivalent in the extreme anisotropic limit to $2K_s(2K_t^\star)^{-1}=
\lambda_c=1$ when equation~(\ref{eq-duality}) is also used.
Fluctuations in the ground state structure have their origin in the 
competition between two non-commuting terms in the Hamiltonian.
In expression~(\ref{eqHITF}), the term 
$-\lambda\hat{\bsigma}_z(j)\hat{\bsigma}_z(j+1)$ acts like a ferromagnetic 
interaction which reinforces order in the $z$-direction (order parameter)
in the chain, while  
$\hat{\bsigma}_x(j)$ appears as a disordering flipping term.
When $\lambda>1$, the ordering term dominates and we expect ferromagnetic
$z$-order in the ground state, 
while the disordered phase corresponds to a dominant role of the flipping term
when $\lambda<1$.

\subsection{The symmetries of the model}
\begin{figure} [th]
\vspace{-1.cm}
        \epsfxsize=7cm
        \begin{center}
        \mbox{\epsfbox{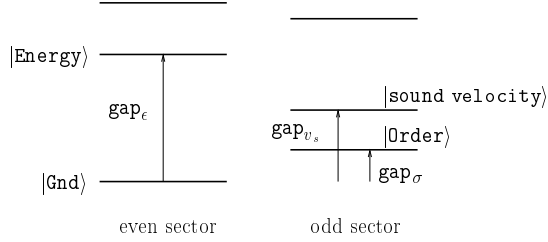}\qquad}
        \end{center}\vskip -6cm
        \caption{Structure of the bottom of the spectrum.}
        \label{Fig2}  
\end{figure}

The classical model exhibits the $Z_2$ symmetry (invariance under global
change $+1\longleftrightarrow -1$ on each site). In the quantum
case we expect a similar
property~\footnote{For simplicity, we use
a short notation $|\uparrow,\downarrow,\uparrow,\dots,\uparrow\>$ for
$|\uparrow_1\>\otimes |\downarrow_2\>\otimes
|\uparrow_3\>\otimes\dots \otimes|\uparrow_L\>$.} 
$|\uparrow,\downarrow,\uparrow,\dots,\uparrow\>_z \longleftrightarrow 
|\downarrow,\uparrow,\downarrow,\dots,\downarrow\>_z$.
This property has an algebraic manifestation through the commutator
\be
	[\mat{H},\mat{P}]=0,\quad\mat{P}=\prod_{j=1}^L\hat{\bsigma}_x(j).
\ee  
The eigenstates of $\mat{H}$ may then be classified according to their parity,
$P=\pm 1$, 
$\mat{P}|{\tt even\ state}\>=+1|{\tt even\ state}\>$ and
$\mat{P}|{\tt odd\ state}\>=-1|{\tt odd\ state}\>$ and an obvious 
consequence is that $\mat{H}$ has vanishing matrix elements between states
of different parities.
More generally, for any {\em even} operator $\mat E$ such that
$\mat E\mat{P}=\mat{P}\mat E$, the only possibly non vanishing matrix 
elements of $\mat E$ are between states of the same parity while
{\em odd} operators satisfying $\mat O\mat{P}=-\mat{P}\mat O$
have surviving matrix elements between states of opposite parities.
The ground state of the system is expected to be even (it is of the
symmetry of the Hamiltonian itself), so that a measure of
local order in the system is given in agreement with equation~(\ref{eq-elmat}) 
by
\be\<{\tt Order\ parameter}\>
=|\<{\tt Lowest\ odd\ state}|\hat{\bsigma}_z(j)
|{\tt Gnd}\>|.\ee
A similar definition of the energy density may be given by inserting 
terms appearing into the Hamiltonian inside states of the same parity,
e.g.
\be\<{\tt Energy\ density}\>=
|\<{\tt 1st\ excited\ even\ state}|\hat{\bsigma}_x(j)|{\tt Gnd}\>|.\ee
Matrix elements of $\hat{\bsigma}_z(j)\hat{\bsigma}_z(j+1)$ might have been
chosen as well.
The reason for being  mainly interested in the bottom of the spectrum
(see figure~\ref{Fig2}) lies in
the fact that the quantum phase transition takes place at zero
temperature.
Together with the eigenstates already mentioned, we define
the state  $|{\tt  sound\ velocity}\>$ and the corresponding
relation which fixes the value of the sound velocity,
\be E_{v_s}-E_{\tt Order}=\frac{2\pi}{L}v_s.\ee
If the normalization is not properly chosen, or if the sound velocity is 
not known, equation~(\ref{eq-gap}) should be replaced by
\be{\tt gap}_\phi=\frac{2\pi}{L}v_s x_\phi\ee
and $v_s$ obtained by the equation above. In the case of 
the quantum Ising model, the Hamiltonian~(\ref{eqHITF}) is conveniently 
normalized and $v_s=1$~\cite{Pfeuty70}.

\subsection{Small Chains}
A quantum chain of length $L$ corresponds to an infinitely long ladder or strip
of classical spins $s_{i,j}$. 
In the case of a chain of 2 spins with periodic boundary conditions (the
classical counterpart is thus a cylinder), the 
Hamiltonian may be written
\be -2\mat{H}=\lambda \hat{\bsigma}_z(1)
	\hat{\bsigma}_z(2)+\lambda \hat{\bsigma}_z(2)
	\hat{\bsigma}_z(1)+
	\hat{\bsigma}_x(1)+
	\hat{\bsigma}_x(2).
\ee
The construction of the matrix of $\mat H$ is easy,
\be
[-2\mat{H}]=
\left(
\begin{array}{cccc}
	2\lambda & 1 & 1 & 0\\
	1 & -2\lambda& 0 & 1\\
	1 & 0 & -2\lambda& 1\\
	0 & 1 & 1 & 2\lambda
\end{array} 
\right)
\begin{array}{c}
|\uparrow,\uparrow\>_z \\ 
|\uparrow,\downarrow\>_z \\
|\downarrow,\uparrow\>_z \\
|\downarrow,\downarrow\>_z
\end{array}
\label{eqMatrixL2}
\ee
but it is not written here in the simplest way.
Due to the parity property of $\mat{H}$, it is easier to write the 
$4\times 4$ matrix in the basis of $\hat{\bsigma}_x$-eigenstates, where it
is block diagonal. In {\em the basis
$\{|\uparrow,\uparrow\>_x , |\downarrow,\downarrow\>_x , 
|\uparrow,\downarrow\>_x ,
|\downarrow,\uparrow\>_x \}$}, we have

\be
[-2\mat{H}]=
\left(
\begin{array}{c|c}
	\begin{array}{cc}
	2& 2\lambda \\
	2\lambda & -2\\
	\end{array}
& \hbox{\LARGE 0} \\
\hline 
\hbox{\LARGE 0} & 
	\begin{array}{cc}
	0& 2\lambda \\
	2\lambda & 0
	\end{array}
\end{array} 
\right)
\begin{array}{c}
|\uparrow,\uparrow\>_x \\ 
|\downarrow,\downarrow\>_x \\
|\uparrow,\downarrow\>_x \\
|\downarrow,\uparrow\>_x
\end{array}
\label{eqMatrixL2}
\ee
In the even sector at the critical coupling $\lambda_c=1$, the eigenvalues
of $\mat H$ are
$E_{\tt Gnd}=-\sqrt{2}$ 
and $E_{\tt Energy}=+\sqrt{2}$. 
The ground state 
$|{\tt Gnd}\>=a_{\tt Gnd}^{\uparrow\uparrow}|\uparrow,\uparrow\>_x
+a_{\tt Gnd}^{\downarrow\downarrow}|\downarrow,\downarrow\>_x$ 
has normalized components $a_{\tt Gnd}^{\uparrow\uparrow}=0.924$ and 
$a_{\tt Gnd}^{\downarrow\downarrow}=0.383$
while the {\em energy} excited state is given by
$|{\tt Energy}\>=a_{\tt Energy}^{\uparrow\uparrow}|\uparrow,\uparrow\>_x
+a_{\tt Energy}^{\downarrow\downarrow}|\downarrow,\downarrow\>_x$ 
with 
$a_{\tt Energy}^{\uparrow\uparrow}=0.383$ and
$a_{\tt Energy}^{\downarrow\downarrow}=-0.924$.
The energy matrix element follows,
\bea \<{\tt Energy\ density}\>&=&
|\<{\tt Energy}|\hat{\bsigma}_x(1)|{\tt Gnd}\>|
\nonumber\\
&=&|a_{\tt Energy}^{\uparrow\uparrow}a_{\tt Gnd}^{\uparrow\uparrow}
-a_{\tt Energy}^{\downarrow\downarrow}a_{\tt Gnd}^{\downarrow\downarrow}|
\nonumber\\
&=&0.707.\eea
In the odd sector, the lowest eigenvalue is 
$E_{\tt Order}=- 1$ 
and the corresponding eigenvector
$|{\tt Order}\>=a_{\tt Order}^{\uparrow\downarrow}|\uparrow,\downarrow\>_x
+a_{\tt Order}^{\downarrow\uparrow}|\downarrow,\uparrow\>_x$ 
has components $a_{\tt Order}^{\uparrow\downarrow}
=a_{\tt Order}^{\downarrow\uparrow}=0.707$. 
The order parameter matrix element follows,
\bea\<{\tt Order\ parameter}\>&=&
|\<{\tt Order}|\hat{\bsigma}_z(1)|{\tt Gnd}\>|
\nonumber\\
&=&|a_{\tt Order}^{\uparrow\downarrow}
a_{\tt Gnd}^{\downarrow\downarrow}+a_{\tt Order}^{\downarrow\uparrow}
a_{\tt Gnd}^{\uparrow\uparrow}|
\nonumber\\
&=&0.924.\eea
In the case of a chain of 3 spins with periodic boundary conditions, 
we have  in the basis of $\sigma_x$ eigenstates 
\be
[-2\mat{H}]=
\left(
\begin{array}{cccc|ccccc}
3 & \lambda & \lambda & \lambda & \\
\lambda & -1 & \lambda & \lambda &  \\
\lambda & \lambda & -1 & \lambda & \\
\lambda & \lambda & \lambda & -1 \\
\hline
&&&& -3 & \lambda & \lambda & \lambda & \\
&&&& \lambda & 1 & \lambda & \lambda & \\
&&&& \lambda & \lambda & 1 & \lambda & \\
&&&& \lambda & \lambda & \lambda & 1
\end{array} 
\right)
\begin{array}{c}
|\uparrow,\uparrow,\uparrow\>_x\\
|\downarrow,\downarrow,\uparrow\>_x\\
|\uparrow,\downarrow,\downarrow\>_x\\
|\downarrow,\uparrow,\downarrow\>_x\\
|\downarrow,\downarrow,\downarrow\>_x\\
|\uparrow,\uparrow,\downarrow\>_x\\
|\downarrow,\uparrow,\uparrow\>_x\\
|\uparrow,\downarrow,\uparrow\>_x
\end{array}
\label{eqMatrixL3}
\ee

The relevant energy levels and corresponding matrix elements
for  three small sizes are collected in 
table~\ref{tab1}. Use will be made of these results to get approximate
values of the critical exponents in the discussion in section~\ref{sec-dissc}.

\begin{table}
        \caption{Values of the physical properties of small quantum
	Ising chains.}
        \label{tab1}
        \begin{tabular}{llllcc}
        \br
	$L$ & $\phantom -E_{\tt Gnd}$ & $\phantom -E_{\tt Energy}$ 
	& $\phantom -E_{\tt Order}$ &
	$\<{\tt Energy\ density}\> $ &
	$\<{\tt Order\ parameter}\> $ \\
        \mr
$2$ & $-1.414$ & $+1.414$ & $-1.000$ &  
$0.707$ & $0.924$ \\
$3$ & $-2.000$ & $\phantom + 0.000$ & $-1.732$ 
	& $0.577$ & $0.880$ \\
$4$ & $-2.613$ & $-1.082$ & $-2.414$  
& $0.462$ & $0.848$ \\
        \br
        \end{tabular}
\end{table}

\section{Quantum Potts chain}
The Potts model generalizes the Ising model. The sites are occupied by
Potts variables (abusively
called spins) with $q$ different states,
$n_{i,j}=0,1,\dots q-1$,  and bonds between nearest neighbour sites
may have  two different energy levels, depending on the relative states of
the site variables, e.g. $-J(q\delta_{n,n'}-1)$.
\subsection{Hamiltonian limit}
We follow the same steps as in the case of the Ising 
model~\cite{SolyomPfeuty81,Hamer81}. 
The energy of a configuration is written as a sum over time slices,
\be\fl
	-\beta E\{n_{i,j}\}=\sum_i\left[
         K_s\sum_{j=1}^L(q\delta_{n_{i,j},n_{i,j+1}}-1)
        +K_t\sum_{j=1}^L(q\delta_{n_{i,j},n_{i+1,j}}-1)\right]
,\label{eq-enPotts}
\ee
where the fact that interactions are limited to  nearest neighbours 
allows a factorized
partition function 
$Z=\Tr_{\{n_{i,j}\}}\prod_i T_{i,i+1}$
with matrix elements between row states
$|\{\nu_j\}\>=|n_{i,1},n_{i,2},\dots n_{i,L}\>$ and 
$|\{{\nu'}_j\}\>=|n_{i+1,1},n_{i+1,2},\dots n_{i+1,L}\>$ at time
indexes $i$ and $i+1$,
\be
T_{i,i+1}=\<\{{\nu'}_j\}|\mat{T}|\{\nu_j\}\>.
\ee 
Now $\mat{T}$ is a $q^{L}\times q^{L}$ matrix
(each of the $L$ spins $n_{i,j}$ have $q$ different possible states).
As in the case of the Ising model, the space coupling term is diagonal, 
$\delta_{\{\nu_j\},\{{\nu'}_j\}}
\times\exp(qK_s\sum_k(\delta_{n_{i,k},n_{i,k+1}}-1))$.
Its description in $\mat T$ requires the introduction of a combination
of  diagonal
operators in the basis $|\{\nu_j\}\>$, the matrix element of which
reproduces the Kronecker delta.
For that purpose, we 
define state vector $|\nu_1\>\otimes |\nu_2\>\otimes\dots\otimes
|\nu_L\>\equiv |\nu_1,\nu_2,\dots,\nu_L\>$ 
and the corresponding
diagonal operators $\mat {C}_k$ and $\mat{C}^\dagger_k$ such 
that (in this section, we stay close to the notations of Ref.~\cite{Turban89}) 
\bea \mat{C}_k|\nu_1,\nu_2,\dots,\nu_k,\dots,\nu_L\>&\equiv&\e^{2i\pi \nu_k/q}
|\nu_1,\nu_2,\dots,\nu_k,\dots,\nu_L\>,\nonumber\\
\mat{C}^\dagger_k|\nu_1,\nu_2,\dots,\nu_k,\dots,\nu_L\>&\equiv&\e^{-2i\pi \nu_k/q}
|\nu_1,\nu_2,\dots,\nu_k,\dots,\nu_L\>.\eea
Due to the property 
$$\frac 1q\sum_{p=0}^{q-1}\e^{-2ip\pi(\nu_k-\nu_l)/q}=\delta_{\nu_k,\nu_l}$$
we may write the Kronecker delta as the diagonal matrix element
\be
	\delta_{\nu_k,\nu_l}=\<\{\nu_j\}|\frac 1q\sum_{p=0}^{q-1}(\mat{C}_k^\dagger
	\mat{C}_l)^p|\{\nu_j\}\>.
\ee
and thus the space contribution to the transfer matrix follows
\be
	\exp\left(K_s\sum_{j=1}^L
	\sum_{p=1}^{q-1}(\mat{C}_j^\dagger
	\mat{C}_{j+1})^p
	\right)
\ee
where the term $p=0$ is subtracted, since it
compensates the $-1$ in the definition of the pair 
energy in equation~(\ref{eq-enPotts}).
For the time contribution, it is necessary to introduce flipping operators
and we define
\bea \mat{R}_k|\nu_1,\nu_2,\dots,\nu_k,\dots, \nu_L\>&\equiv&
|\nu_1,\nu_2,\dots,\nu_{k}+1,\dots ,\nu_L\>,\nonumber\\
\mat{R}^\dagger_k|\nu_1,\nu_2,\dots,\nu_k,\dots ,\nu_L\>&\equiv&
|\nu_1,\nu_2,\dots,\nu_{k}-1,\dots ,\nu_L\>,
\eea
where periodicity in space state is assumed, $|\nu+q\>=|\nu\>$.
The matrix representation of these operators is the following
\be\fl
[\mat{C}_j]=\mat 1\otimes\mat 1\otimes\dots\otimes
\left(
\begin{array}{ccccc}
1 & 0 & 0& \dots  & 0\\
0 & \e^{2i\pi/q} & 0&\dots  & 0 \\
\vdots & 0 & \e^{4i\pi/q} &\ddots &\vdots\\
0& \vdots & \ddots &\ddots& 0\\
0&0&\dots  & 0& \e^{2i\pi(q-1)/q}
\end{array}
\right)_j
\dots\otimes\mat 1,
\ee
\be\fl
[\mat{R}_j]=\mat 1\otimes\mat 1\otimes\dots\otimes
\left(
\begin{array}{ccccc}
0 & 0 & \dots  &0& 1\\
1 & 0 & \dots  & 0& 0 \\
0& 1 & \ddots & \vdots & \vdots\\
\vdots &\ddots&1&0&0\\
0&\dots  &0& 1& 0
\end{array}
\right)_j
\dots\otimes\mat 1.
\ee
They commute on different sites and obey the following algebra on a given
site
$\mat{R}_j\mat{C}_j=\e^{-2i\pi/q}\mat{C}_j\mat{R}_j$,
$\mat{R}^\dagger_j\mat{C}_j=\e^{2i\pi/q}\mat{C}_j\mat{R}^\dagger_j$, 
$\mat{R}_j\mat{C}^\dagger_j=\e^{2i\pi/q}\mat{C}^\dagger_j\mat{R}_j$,
$\mat{R}^\dagger_j\mat{C}^\dagger_j=
\e^{-2i\pi/q}\mat{C}^\dagger_j\mat{R}^\dagger_j$, 
$\mat{C}_j^q=
\mat{R}_j^q=\mat{1}$.
Temporarily forgetting 
about the site index, diagonalization of $\mat{R}$ leads to
$$\mat{R}|r\>=\e^{2ir\pi/q}|r\>,\quad |r\>=
q^{-1/2}\sum_{\nu=0}^{q-1}\e^{-(2ir\pi/q)\nu}
|\nu\>.$$
It follows that the operator $\hat{\bTheta}\equiv
\frac 1q\sum_{p=0}^{q-1}\mat {R}^p$ 
has the property 
$\hat{\bTheta}|r\>=\delta_{r,0}|r\>$
(it is equal the projector
on the ``zero eigenstate'' of $\mat R$,
 $\hat{\bTheta}=|r=0\>\<r=0|$,
and so $\<\{{\nu'}_j\}|\hat{\bTheta}|\{\nu_j\}\>=1/q$
for any pair of states $|\{\nu_j\}\>$ and $|\{{\nu'}_j\}\>$) which enables to write
\be
	\exp(qK_t^\star\hat{\bTheta}_j)=\mat 1+\hat{\bTheta}_j
	(\e^{qK_t^\star}-1)
\ee
for arbitrary value of $K_t^\star$.
The time contributions, $\e^{K_t(q\delta_{\nu_j,{\nu'}_j}-1)}$, which can 
take values $\e^{K_t(q-1)}$ if ${\nu'}_j=\nu_j$ and $\e^{-K_t}$ otherwise,
are obtained from matrix elements of terms 
$\exp(qK_t^\star \hat{\bTheta}_j)$
provided that $K_t^\star$ satisfies the duality relation
\be
(\e^{qK_t}-1)(\e^{qK_t^\star}-1)=q.
\ee
The transfer matrix eventually reads as
\be
	\mat T=	\exp\left(K_s\sum_{j=1}^L
	\sum_{p=1}^{q-1}(\mat{C}_j^\dagger
	\mat{C}_{j+1})^p
	\right)
	\times
	\exp\left(K_t^\star\sum_{j=1}^L
	\sum_{p=1}^{q-1}\mat{R}_j^p
	\right).
\ee 
We have shifted the sum in the last term, starting from $p=1$ which only
changes the transfer matrix by a constant prefactor $\e^{K_t^\star}$.
This modification affects the free energy density (or the Hamiltonian)
by a constant only and does not change the thermodynamic
properties.
In the Hamiltonian limit $K_s\to 0$, $K_t\to\infty$ with fixed $\lambda=
K_s/K_t^\star$, we obtain the Hamiltonian of the quantum Potts chain 
\be
	\mat H=	-\frac 12\lambda\sum_{j=1}^L
	\sum_{p=1}^{q-1}(\mat{C}_j^\dagger
	\mat{C}_{j+1})^p
	-\frac 12\sum_{j=1}^L
	\sum_{p=1}^{q-1}\mat{R}_j^p
	.
\ee 
It is easy to check that in the case $q=2$, we 
recover the Hamiltonian~(\ref{eqHITF}) of the quantum Ising chain with
$\mat C$ playing the role of $\hat{\bsigma}_z$
and $\mat R$ that of $\hat{\bsigma}_x$. The limit $\lambda\to\infty$
leads to $q$ degenerate ordered ground states $|n,n,n,\dots,n\>$,
$n=0,1,\dots, q-1$, while in the other limit $\lambda\to 0$, the term in
$\mat{R}_j$  introduces disorder in the
ground state through local rotations
between $\mat{C}_j$-eigenstates.
The critical point of the classical two-dimensional model is given by duality,
$(\e^{qK_s}-1)(\e^{qK_t}-1)=q$, i.e. $K_s=K_t^\star$ or $\lambda_c=1$.

A different, but simpler route in order to get the time contribution to the
Hamiltonian matrix is the 
following: 
in the ``row states basis'' $|\{\nu_j\}\>$,  
the time contribution
$
	\prod_k\e^{K_t(q\delta_{\nu_k,{\nu'}_{k}}-1)}
$
is a product over ``single bond time transfer operators'' 
\be
\fl	{\tt time\ contribution\ to\ \mat T}\equiv
	\prod_k{\sum}_{\{\nu_j\},\{{\nu'}_j\}}
	|\{{\nu'}_j\}\>\e^{K_t(q\delta_{\nu_k,
	{\nu'}_k}-1)} \<\{\nu_j\}|
\ee
which take
values $\e^{K_t(q-1)}$ on the diagonal and the same value $\e^{-K_t}$
for all single spin flipping terms, $\e^{-2K_t}$
for simultaneous double spin flipping terms and so on. Anticipating
further simplifications which occur while taking the Hamiltonian limit
$K_t\to\infty$, we  restrict ourselves to single flipping terms.
We introduce the symbol ${\sum}'_{\{\nu_j\},\{{\nu'}_j\}}$ 
with the meaning of a double 
sum over row states $|\{\nu_j\}\>$ and $|\{{\nu'}_j\}\>$ such that
$\forall j\not=k$, ${\nu'}_j=\nu_j$ (single spin flipping terms, since only
$\nu_k$ is likely to be flipped between the two row states).
The contribution to the Hamiltonian follows from taking the logarithm of this
expression which leads to $-\epsilon\mat H$.
\bea\fl
\ln ({\tt time\ contribution\ to\ \mat T})
	&=&
	\sum_k\ln\Bigl[\e^{K_t(q-1)}
		\bigl(
		\mat 1+\nonumber\\
	&&\e^{-qK_t}
	{\sum}'_{{\{\nu_j\},\{{\nu'}_j\}\atop {\nu'}_k\not =\nu_k}}
	|\{{\nu'}_j\}\> 
	\<\{\nu_j\}|\bigr)+O(\e^{-(q+1)K_t})\Bigr]\nonumber\\
	&\simeq&LK_t(q-1)+\e^{-qK_t}
	{\sum}'_{{\{\nu_j\},\{{\nu'}_j\}\atop {\nu'}_k\not =\nu_k}}
	|\{{\nu'}_j\}\> 
	\<\{\nu_j\}|
\eea
in the limit of a strong coupling in the time direction, $K_t\to\infty$.

\subsection{Symmetries}
We proceed as in the Ising case to get the symmetries of the Hamiltonian.
The classical Potts model is obviously globally unchanged by the 
cyclic transformation $\forall j$, $0_j\to 1_j$, $1_j\to 2_j$, \dots
$(q-1)_j\to 0_j$. Such a rotation in spin states is realized by the operator
$\mat R_j$. We may thus define a charge operator which simultaneously
rotates Potts variables at all sites, 
\be\mat Q=\prod_j\mat R_j.\ee
 Using the commutation relations between the $\mat R_j$'s and the
$\mat C_j$'s, it is easy to prove that the charge operator commutes with
the Hamiltonian. Since the eigenvalues of $\mat R_j$ are the $q$ distinct 
complex numbers
$\omega_j=\e^{2i\pi r_j/q}$, those of the charge operator are also given by
$q$ numbers $\exp(\sum_j 2i\pi r_j/q)$ which enable to write the Hamiltonian
matrix under a block-diagonal structure with $q$ different sectors, depending
on the value of $\sum_j r_j\ {\rm mod}(q)$ (the sectors can be referred to as
sector \# $0$, $1$, \dots, $q-1$).

Any operator $\mat O_p$  with the commutation property
$\mat O_p\mat Q=\e^{2i\pi p/q}\mat Q\mat O_p$ ($\mat R$ is such an example
with $p=0$, $\mat C^\dagger$ is an example with $p=1$, and $\mat C$ with 
$p=q-1$) has non vanishing matrix elements between states of defined 
symmetry provided that the charge sectors obey a simple relation
$\nu_\psi-\nu_\phi= p$:
\bea
	&&\forall |\phi\>,|\psi\>\ {\rm such\ that}\  
	\mat Q|\phi\>=\e^{2i\pi \nu_\phi/q}|\phi\>,
	\mat Q|\psi\>=\e^{2i\pi \nu_\psi/q}|\psi\>,\ \nonumber\\
	&&\nu_\psi-\nu_\phi\not= p\Rightarrow\<\phi|\mat O_p|\psi\>=0
\eea
As an energy density, we may choose any single term appearing inside the 
Hamiltonian, 
e.g. ${\tt Energy\ density}=\mat R_j$. 
A measure of the order parameter would be
provided by ${\tt Local\ order\ parameter}=\mat C_j$, or $\mat C_j^\dagger$ as 
well. 
The energy density gives access to the critical exponent $x_\epsilon$.
For the order parameter, although the two choices 
correspond to different values of $p$, since two 
critical exponents are sufficient in order to determine the whole 
universality class, the same value of $x_\sigma$ is expected from both
definitions. From gap scaling in particular, we expect $(q-1)$-fold degeneracy
of the sectors $p\not=0$.

\subsection{Diagonalization of small chains}
In the case of a quantum $3-$state 
Potts chain of length $L=2$ with periodic boundary conditions, 
the action of the Hamiltonian in the $\mat C-$operators eigenbasis
is the following,
\bea &-&2\mat H|\nu_1,\nu_2\>=2
	\lambda(\omega_1\omega_2^{-1}+\omega_1^2\omega_2^{-2})
|\nu_1,\nu_2\>\nonumber\\
	&&\qquad+|\nu_1+1,\nu_2\>
	+|\nu_1+2,\nu_2\>+|\nu_1,\nu_2+1\>+|\nu_1,\nu_2+2\>,\\
&&\omega_j=\exp(2i\pi \nu_j/3).
\eea
We obtain the following
$3^2\times 3^2$ matrix,
\be
	\fl[-2\mat H]=\left(
	\begin{array}{ccccccccc}
	4\lambda & 1 & 1 & 1 & 0 & 0 & 1 & 0 & 0 \\
	1&-2\lambda&1&0&1&0&0&1&0\\
	1&1&-2\lambda&0&0&1&0&0&1\\
	1&0&0&-2\lambda&1&1&1&0&0\\
	0&1&0&1&4\lambda&1&0&1&0\\
	0&0&1&1&1&-2\lambda&0&0&1\\
	1&0&0&1&0&0&-2\lambda&1&1\\
	0&1&0&0&1&0&1&-2\lambda&1\\
	0&0&1&0&0&1&1&1&4\lambda\\
	\end{array}
	\right)
	\begin{array}{c}
	|00\>_{C}\\|01\>_{C}\\|02\>_{C}\\|10\>_{C}\\|11\>_{C}\\|12\>_{C}
	\\|20\>_{C}\\|21\>_{C}\\|22\>_{C}\\
	\end{array}
\ee
the eigenvalues of which are (eigenvalues of $\mat H$ at $\lambda_c=1$), 
$-2.732$, $-2.137$, $-2.137$, $+0.500$, $+0.500$, $+0.732$, $+1.637$, $+1.637$,
and $+2.000$.
It is of course more efficient to exploit the symmetries and to work in the 
$\mat R-$operators eigenbasis
where,
\bea &-&2\mat H|r_1,r_2\>=2
	\lambda(|r_1+1,r_2-1\>+|r_1+2,r_2-2\>)\nonumber\\
	&&\qquad+(\omega_1+\omega_1^2+\omega_2+\omega_2^2)|r_1,r_2\>,
	\\
&&\omega_j=\exp(2i\pi r_j/3).
\eea
The corresponding matrix is now
\be\fl
[-2\mat H]=\left(
\begin{array}{ccc}
\begin{array}{ccc|}
4 & 2\lambda & 2\lambda \\
2\lambda & -2 & 2\lambda \\
2\lambda & 2\lambda & -2 \\
\hline
\end{array}\\
&\begin{array}{|ccc|}
\hline
-2 & 2\lambda & 2\lambda \\
 2\lambda & 1 & 2\lambda \\
 2\lambda & 2\lambda & 1 \\
\hline
\end{array}\\
&&\begin{array}{|ccc}
\hline
-2 & 2\lambda & 2\lambda \\
 2\lambda & 1 & 2\lambda \\
 2\lambda & 2\lambda & 1 \\
\end{array}\\
\end{array}\right)
\begin{array}{c}
	|00\>_{R}\\|12\>_{R}\\|21\>_{R}\\|11\>_{R}\\|20\>_{R}\\|02\>_{R}
	\\|22\>_{R}\\|01\>_{R}\\|10\>_{R}\\
\end{array}
\ee
It is block diagonal, with eigenvalues of the first block ($0$ sector) 
$-2.732$, $+0.732$ 
and $+2.000$, and the two-fold degenerate eigenvalues of the two remaining
identical blocks (sectors $1$ and $2$), $ -2.137$, $+0.500$ and $+1.637$.
The ground state and energy eigenstates are given by
$|{\tt Gnd}\>=a^{00}_{\tt Gnd}|00\>+
a^{12}_{\tt Gnd}|12\>+a^{21}_{\tt Gnd}|21\>$ and 
$|{\tt Energy}\>=a^{00}_{\tt Energy}|00\>+
a^{12}_{\tt Energy}|12\>+a^{21}_{\tt Energy}|21\>$ 
and the energy density follows
\bea \<{\tt Energy\ density}\> &=&
|\<{\tt Energy}|\mat R_1|{\tt Gnd}\>|
\nonumber\\
&=&|a^{00}_{\tt Energy}a^{00}_{\tt Gnd}
+a^{12}_{\tt Energy}a^{12}_{\tt Gnd}\e^{2i\pi/3}
+a^{21}_{\tt Energy}a^{21}_{\tt Gnd}\e^{4i\pi/3}|
\nonumber\\
&=&0.613.\eea
In the $p=1$ sector, the order parameter density is given by (the 
notation is obvious)
\bea \<{\tt Order\ parameter}\> &=&
|\<{\tt Order}\ p=1|\mat C^\dagger_1|{\tt Gnd}\>|
\nonumber\\
&=&|a^{10}_{\tt Order}a^{00}_{\tt Gnd}
+a^{22}_{\tt Order}a^{12}_{\tt Gnd}
+a^{01}_{\tt Order}a^{21}_{\tt Gnd}|
\nonumber\\
&=&0.916,\eea
and, due to the complete degeneracy of the two sectors $p=1$ and $p=2$, 
we get the same result in the remaining sector
\bea \<{\tt Order\ parameter}\> &=&
|\<{\tt Order}\ p=2|\mat C_1|{\tt Gnd}\>|
\nonumber\\
&=&|a^{20}_{\tt Order}a^{00}_{\tt Gnd}
+a^{02}_{\tt Order}a^{12}_{\tt Gnd}
+a^{11}_{\tt Order}a^{21}_{\tt Gnd}|
\nonumber\\
&=&0.916,\eea

The relevant eigenvalues and matrix elements
 for systems of sizes $L=2$ and 3
 are collected in table~\ref{tabPotts1}.

\begin{table}
        \caption{Values of the physical properties of small quantum
	$3-$ (upper part of the table) and $4-$state (lower part of the table)
	Potts chains.}
        \label{tabPotts1}
        \begin{tabular}{llllcc}
        \br
	$L$ & $\phantom -E_{\tt Gnd}$ & $\phantom -E_{\tt Energy}$ & 
	$\phantom -E_{\tt Order}$ &
	$\<{\tt Energy\ density}\> $ &
	$\<{\tt Order\ parameter}\> $ \\
        \mr
$2$ & $-2.732$ & $\phantom -0.732$ & $-2.137$ 
	& $\phantom -0.613$ & $\phantom -0.916$ \\
$3$ & $-3.842$ & $-1.500$ & $-3.462$
	& $\phantom -0.505$ 
	& $\phantom -0.867$\\
        \br
$2$ & $-4.000$ & $\phantom -0.000$ & $-3.236$ & $\phantom -0.577$ 
	& $\phantom -0.909$ \\
$3$ & $-5.606$ & $-3.000$ & $-5.123$ & $\phantom - 0.480$ & $\phantom -0.858$\\
\br
        \end{tabular}
\end{table}

Interested readers might find helpful 
to have also the matrix representing the $4-$state Potts chain. When $L=2$
(the matrix is now $4^L\times 4^L$)
at the critical coupling $\lambda_c=1$  
in its block diagonal form in the $\mat R-$eigenbasis
$\{|00\> , |13\> , |22\> , |31\> , |01\> , |10\> 
	, |23\> , |32\> , |02\> , 
	|11\> , |20\> , |33\> , |03\> , |12\> , |21\> , 
	|30\>  \}$, this matrix reads as
\be\fl\scriptsize
[-2\mat H]=\left(
\begin{array}{cccc}
\begin{array}{rrrr|}
6 & 2  & 2  & 2 \\
2  & -2 & 2  & 2  \\
2  & 2  & -2 & 2  \\
2  & 2  & 2  & -2\\
\hline
\end{array}\\
&\begin{array}{|rrrr|}
\hline
2 & 2  & 2  & 2  \\
 2  & 2 & 2  & 2  \\
 2  & 2  & -2 & 2  \\
 2  & 2  & 2  & -2\\
\hline
\end{array}\\
&&\begin{array}{|rrrr|}
\hline
2 & 2  & 2  & 2  \\
 2  & 2 & 2  & 2  \\
 2  & 2  & -2 & 2  \\
 2  & 2  & 2  & -2\\
\hline
\end{array}\\
&&&\begin{array}{|rrrr}
\hline
2 & 2  & 2  & 2  \\
 2  & 2 & 2  & 2  \\
 2  & 2  & -2 & 2  \\
 2  & 2  & 2  & -2\\
\end{array}\\
\end{array}\right)
\ee

The following section, where an approximate determination of
the critical exponents of the $3-$ and $4-$state Potts  models 
will be proposed, will make use of the numerical values listed in 
table~\ref{tabPotts1}.

\section{Discussion}\label{sec-dissc}
Finite-size estimators of the critical exponents follow from 
equations~(\ref{eq-e}) and (\ref{eq-m}),
\be x_\phi=\frac{\ln\phi(L)-\ln\phi(L')}{\ln L-\ln L'}.\ee 
The results, collected
in table~\ref{tab2} in the case of the Ising, $3-$ and $4-$state Potts models,
show as expected a very weak convergence. A strip of width $2$ to $4$ is 
obviously a poor approximation of the thermodynamic limit.
In the case of the Ising model, exact 
diagonalization~\cite{LiebEtAl61,Pfeuty70} 
through Jordan-Wigner transformation into
fermion operators, then Bogoljubov-Valatin canonical transformation
into free fermions leads to an exact expression 
for the energy density~\cite{TurbanBerche93}
\be\<{\tt Energy\ density}\>=\frac 2L\cos\left(\frac\pi{2L}\right).
\ee
Unfortunately, there is no closed expression for the order parameter
matrix element which couples two different sectors of the Hamiltonian. 
Hence,  the presence of a boundary term breaks the
quadratic expression necessary for the diagonalization. 

\begin{table}
        \caption{Estimators of critical exponents from FSS of local
	properties of small quantum
	Ising and $3-$ and $4-$state Potts chains.}
        \label{tab2}
        \begin{tabular}{llllllllllll}
        \br
	\centre{4}{Ising}&\centre{4}{$3-$state Potts}&
	\centre{4}{$4-$state Potts}\\
	\crule{4}&\crule{4}&\crule{4}\\
	$L$ & $L'$ & $x_\epsilon$ & $x_\sigma$ 
	&$L$ & $L'$ & $x_\epsilon$ & $x_\sigma$
	&$L$ & $L'$ & $x_\epsilon$ & $x_\sigma$ 
	 \\
\mr
2 & 3 & 0.501 & 0.120 & 2 & 3 & 0.478 & 0.136 & 2 & 3 & 0.454 & 0.142\\
3 & 4 & 0.773 & 0.129 &  -- & -- &  --     &  -- & -- & -- & -- & --    \\
\mr
$\infty$ & & 1.000 & 0.125 & $\infty$ & & 0.800 & 0.133
& $\infty$ & & 0.500 & 0.125 \\
        \br
        \end{tabular}
\end{table}

The resort to 
gap scaling~(\ref{eq-gap}) 
as predicted from conformal invariance is thus desirable.
Indeed, as emphasized in Section~2, the cylinder geometry is
the geometrical shape of the classical problem corresponding to the
one-dimensional quantum chain. Since such an infinitely long cylinder
follows from the infinite plane geometry through conformal mapping, 
we may argue that the thermodynamic limit is somehow encoded in the
results following from conformal rescaling and a better convergence
for the critical exponents is expected.
With the normalization of equation~(\ref{eqHITF}) the fermion excitations
take the form $\varepsilon_k=|2\sin\frac k2|$. The sound velocity is thus
fixed to unity as previously announced.
The  results following from gap scaling,
\be x_\phi=\frac{L\times{\tt gap}_\phi}{2\pi v_s},\ee 
are collected in table~\ref{tab3}. The quality of the results is indisputably
better than through FSS.

In the case of the Potts model, the sound velocity is not known
and it is worth referring to the literature. Gehlen et al. have studied 
numerically quantum
Potts chains in refs.~\cite{GehlenEtAl84,GehlenEtAl85} where tables of 
numerical results 
are reported for $q=3$ up to a size $L=13$ and for $q=4$ 
up to $L=11$ (they denote $R$ ($P$) the energy gap
(magnetic gap) multiplied by the strip size) . 
These authors used a different 
normalization and their Hamiltonian for $q=3$ in~\cite{GehlenEtAl84}  
is related to ours through
$\frac 23\mat H_{\tt here}(q=3)+\frac 43\mat 1=
\mat H_{\tt Gehlen}(q=3)$, while 
in~\cite{GehlenEtAl85} it is related to ours through  
$\mat H_{\tt here}(q=4)=2\mat H_{\tt Gehlen}(q=4)$. Using the 
values of the sound velocity quoted in~\cite{GehlenEtAl85}, we deduce
that in our case the sound velocity is $3/2$ 
larger for $q=3$ and $2$ times larger
for $q=4$, i.e.
\be v_s(q=3)=1.299,\quad v_s(q=4)=1.578.\ee
These values are used to calculate the exponents reported in 
table~\ref{tab3}. Further numerical values denoted with an asterisk are taken
from 	refs.~\cite{GehlenEtAl84,GehlenEtAl85} to complete the table 
with results at larger sizes. The expected result in the thermodynamic
is also mentioned in the table~\cite{Wu82}.
\begin{table}
        \caption{Estimators of critical exponents from gap scaling 
	of small quantum
	Ising, $3-$state and $4-$state Potts chains. An asterisk indicates that
	the numerical value was extracted from 
	refs.~\cite{GehlenEtAl84,GehlenEtAl85} according to the transformation
	mentioned above, $x=3R/4\pi v_s$ for $q=3$ and $x=R/\pi v_s$ 
	for $q=4$.}
        \label{tab3}
        \begin{tabular}{lllllllll}
        \br
	\centre{3}{Ising}&\centre{3}{$3-$state Potts}&
	\centre{3}{$4-$state Potts}\\
	\crule{3}&\crule{3}&\crule{3}\\
	$L$ & $x_\epsilon$ & $x_\sigma$
	&	$L$ & $x_\epsilon$ & $x_\sigma$
	&	$L$ & $x_\epsilon$ & $x_\sigma$ \\
\mr
2  & 0.900 & 0.132 & 2  & 0.848 & 0.145 & 2 & 0.807 & 0.154 \\
3  & 0.955 & 0.128 & 3  & 0.861 & 0.139 & 3 & 0.788 & 0.146 \\
4  & 0.975 & 0.127 & 4  & 0.857 & 0.137 
	& 4 & 0.767$^\star$ & 0.143$^\star$\\
&&&13 & 0.848$^\star$ & 0.134$^\star$  
	& 11 & 0.700$^\star$ & 0.137$^\star$ \\
\mr
$\infty$ & 1.000 & 0.125& $\infty$ & 0.800 & 0.133 & $\infty$ & 0.500 & 0.125\\
\br
        \end{tabular}
\end{table}

To summarize, we note that the quality of the results is 
acceptable for the relatively 
small effort of diagonalization of small matrices. The quest for 
critical exponents is an important step in the characterization of the nature
of a phase transition, since these quantities are universal. The study of
models of statistical physics which display second-order phase transitions
usually requires sophisticated methods, and approximate determinations
are often desirable. 
Considering quite small systems, we reach a few percent accuracy in 
the determination of the critical 
exponents in the case of the Ising and $3-$state Potts
models. The convergence is poor in the $4-$state Potts case. This is 
essentially due to the logarithmic corrections present in this model.

Eventually, regarding the relatively small efforts, 
we believe that an introduction to the study of quantum chains
in courses on statistical physics or many-body problems might be of interest.


\end{document}